\documentclass[aip, apl, amsmath, amssymb, reprint, footinbib,%
]{revtex4-1}

\usepackage{graphicx}
\graphicspath{{Figures/}}
\usepackage{dcolumn}
\usepackage{bm}

\usepackage[utf8]{inputenc}
\usepackage[T1]{fontenc}
\usepackage{mathptmx}
\usepackage{etoolbox}
\usepackage[dvipsnames]{xcolor}
\usepackage{hyperref}
\usepackage{xfrac}
\usepackage{mathtools}
\usepackage{euscript}
\usepackage{xifthen}

\usepackage[textwidth=1.05cm, textsize=scriptsize]{todonotes}
\usepackage{ulem}
\usepackage{cancel}

\makeatletter
\def\@email#1#2{%
 \endgroup
 \patchcmd{\titleblock@produce}
  {\frontmatter@RRAPformat}
  {\frontmatter@RRAPformat{\produce@RRAP{*#1\href{mailto:#2}{#2}}}\frontmatter@RRAPformat}
  {}{}
}%
\makeatother

\begin{document}
  \newcommand\numberthis{\addtocounter{equation}{1}\tag{\theequation}}

\newcommand{\operator}[3]{\widehat{#1}_{#2}^{#3}}
\newcommand{\phsymbol}{a} 
\newcommand{\xsymbol}{X} 
\newcommand{\plsymbol}{P} 
\newcommand{\phcr}{\operator{\phsymbol}{ln}{\dagger}}
\newcommand{\phan}{\operator{\phsymbol}{ln}{}}
\newcommand{\xcr}{\operator{\xsymbol}{\lambda\nu}{\dagger}}
\newcommand{\xan}{\operator{\xsymbol}{\lambda\nu}{}}
\newcommand{\xcrl}{\operator{\xsymbol}{l \nu}{\dagger}}
\newcommand{\xanl}{\operator{\xsymbol}{l \nu}{}}
\newcommand{\plcr}{\operator{\plsymbol}{ln}{\dagger}}
\newcommand{\plan}{\operator{\plsymbol}{ln}{}}
\newcommand{\x}{\widehat{\xsymbol}}
\newcommand{\pl}{\widehat{\plsymbol}}
\newcommand{\ph}{\widehat{\phsymbol}}
\newcommand{\ham}{\operator{H}{\text{tot}}{}}
\newcommand{\hx}{\operator{H}{\text{x}}{}}
\newcommand{\hp}{\operator{H}{\text{p}}{}}
\newcommand{\hpx}{\operator{H}{\text{xp}}{}}
\newcommand{\hxx}{\operator{H}{\text{xx}}{}}
\newcommand{\hpp}{\operator{H}{\text{pp}}{}}
\newcommand{\gllnn}{g_{ln\lambda\nu}}
\newcommand{\lmnu}{{\lambda\nu}}
\newcommand{\hc}{\text{h.c.}}
\newcommand{\abohr}{a_{\text{Bohr}}}
\newcommand{\auc}{a_{\text{u.c.}}}
\newcommand{\const}{\mathrm{const}}
\newcommand{\glnn}{f_{l\nu n}}
\newcommand{\vdcv}{\vec{d}_{\text{cv}}}
\newcommand{\dcv}{d_{\text{cv}}}
\newcommand{\vint}{V_{\text{int}}}
\newcommand{\vintln}{V_{\text{int}, ln}}
\newcommand{\vintxx}{V_{\text{int,xx}}}
\newcommand{\vintc}{V_{\text{C}}}
\newcommand{\erho}{\vec{e}_{r}}
\newcommand{\etheta}{\vec{e}_{\theta}}
\newcommand{\Eg}{E_{\text{g}}}
\newcommand{\estatic}{\epsilon_{\text{st}}}
\newcommand{\hence}{\, \Rightarrow \, }
\newcommand{\hOmegaRS}{\hbar\Omega_{2\text{D}}}
\newcommand{\hOmegax}{\hbar\Omega_{\text{x}}}
\newcommand{\dtau}{\diff^8 \vec{\tau}}
\newcommand{\Aex}{\EuScript{A}}
\newcommand{\Bex}{\EuScript{B}}
\newcommand{\Cph}{\EuScript{C}}
\newcommand{\eV}{\text{eV}}
\newcommand{\meV}{\text{meV}}
\newcommand{\nm}{\text{nm}}
\newcommand{\kappall}{\kappa_1 \kappa_2 \kappa_3 \kappa_4}
\newcommand{\kappallprime}{\kappa'_1 \kappa'_2 \kappa'_3 \kappa'_4}
\newcommand{\varkappall}{\varkappa_1 \varkappa_2 \varkappa_3 \varkappa_4}
\newcommand{\pump}{\EuScript{S}}
\newcommand{\ep}[1][]{
\ifthenelse{\isempty{#1}}{\vec{e}_{\text{p}}}{e_{\text{p},{#1}}}
}

\newcommand{\vll}{V_{\kappall}}
\newcommand{\suml}{\sum\limits_{l=-\infty}^{+\infty}}
\newcommand{\sumln}{\sum\limits_{l, n}}
\newcommand{\sumn}{\sum\limits_{n=1}^{N(l)}}
\newcommand{\sumnu}{\sum\limits_{\nu=1}^{+\infty}}
\newcommand{\sumnnu}{{\sum\limits_{n, \nu}}^{\prime}}
\newcommand{\sumlmnu}{\sum\limits_{\lambda, \nu}}
\newcommand{\parenth}[1]{\left( {#1} \right)}
\newcommand{\brackets}[1]{\left[ {#1} \right]}
\newcommand{\braces}[1]{\left\{ {#1} \right\}}
\newcommand{\vr}{\vec{r}}
\newcommand{\vrho}{\vec{\rho}}
\newcommand{\homegaln}{\hbar\omega_{ln}}
\newcommand{\homega}{\hbar\omega}
\newcommand{\hOmegaln}{\hbar\Omega_{\lmnu}}
\newcommand{\hOmega}{\hbar\Omega}
\renewcommand{\vec}[1]{\boldsymbol{#1}}
\newcommand{\diag}[1]{\mathop{\mathrm{diag}}\nolimits{#1}}
\newcommand{\abs}[1]{\left| {#1} \right|}
\newcommand{\diff}{\mathrm{d}}

\newcommand{\e}{\varepsilon}
\newcommand{\kz}{\textsl{\ae}}
\newcommand{\elec}{\text{e}}
\newcommand{\hole}{\text{h}}
\newcommand{\mh}{m_{\hole}}
\newcommand{\me}{m_{\elec}}


\newcommand{\inside}{^{(1)}}
\newcommand{\outside}{^{(2)}}
\newcommand{\etal}{\textit{et al.}}
\newcommand{\ptf}{\mathcal{F}_{\text{ptf}}}
\newcommand{\fx}{\text{fx}}
\newcommand{\xx}{\text{xx}}
\newcommand{\dir}{\text{dir}}
\newcommand{\sumprime}{\mathop{{\sum}'}}
\newcommand{\mum}{\mu\text{m}}
\newcommand{\ee}{\mathrm{e}}

  \newcommand{\yuk}[1]{\textcolor{Red}{#1}}

  \preprint{AIP/123-QED}

  \title{Interacting exciton-polaritons in cylindric micropillars}
  \author{Yury S. Krivosenko}
    \email{y.krivosenko@gmail.com}
  \author{Ivan V. Iorsh}
  \author{Ivan A. Shelykh}
    \altaffiliation{Science Institute, University of Iceland, Dunhagi 3, IS-107, Reykjavik, Iceland}
  \affiliation{Department of Physics and Engineering, ITMO University, St.\,Petersburg 197101, Russia}

  \begin{abstract}
    We present a quantitative microscopic analysis of the formation of exciton-polaritons, the composite particles possessing light and material components, polariton-polariton interactions, and resonant pumping dynamics in cylindrical semiconductor micropillars. We discuss how the redistribution effect can be used in devices generating photons with non zero orbital angular momenta.
  \end{abstract}
  \maketitle
  \allowdisplaybreaks

  \begin{section}{Introduction \label{sec:Intro}}
    Cavity polaritons, also known as  exciton-polaritons\cite{Kavokin2017_OxfPr, Ciuti2013_RMP, Deng2010_RMP} are composite quasiparticles consisting of  excitonic and a photonic components. Conventional geometry where polaritons can be routinely observed is a planar semiconductor microcavity, which consists of a semiconductor quantum well, sandwiched between a pair of distributed Bragg reflectors (DBRs) and placed in the position where photonic mode of such Fabry-Perot cavity has an antinode.

    The physics of polaritons attracted substantial interest of the researches working in the domains of photonics and condensed matter physics. This is mainly due to the unique properties of polaritons, related to their composite nature. The combination of extremely low effective mass, inherited from the photonic component, with giant nonlinear optical response, stemming from the excitonic part, enables the observation of a set of intriguing collective phenomena at surprisingly high temperatures \cite{Ciuti2013_RMP,ByrnesRev,Claude2022}. Examples include polariton BEC and polariton lasing \cite{Kasprzak2006,Balili2007,Schneider2013,Riminucci2022}, formation of topological defects such as solitons \cite{Amo2011,Chana2015,Sich2018,Pernet2022_NatPhys}, quantized vortices \cite{Gao2018,Kwon2019} and vortex lattices \cite{Panico2021,Sitnik2022}, skyrmions and merons \cite{Bieganska2021,Lei2022}, shock waves \cite{Bienaime2021} and many others. Moreover, polariton systems can form a basis for creation of nanophotonic devices of the next generation, including optical logic gates and all-optical integrated circuits \cite{Amo2010_NatPhot, Shelykh2011, Askitopoulos2018, Chen2022}.

    The visible current trend in polaritonics is the shift towards quantum applications \cite{Matutano2019,Zasedatelev2021,Nigro2022}. Geometries with cavity polaritons were suggested as candidates for creation of the sources of single photons \cite{Verger2006_PRB,Liew2010,Snijders2018,Delteil2019_NatMat,Kyriienko2020} and entangled photon pairs \cite{Johne2008,Cuevas2018,Denning2022}, the corresponding onset of polaritonic quantum correlations was reported \cite{Trivedi2019,Levinsen2019,Autry2020,Wang2021}. Moreover, the possibility to use polariton systems in quantum computing \cite{Ghosh2020_QI,Kavokin2022} and as quantum and neuromorphic simulators was actively discussed  \cite{Kavokin2022,Kalinin2018,Suchomel2018,Opala2019,Boulier2020_AQT,Marcucci2020}.

    For applications of polaritonics at the quantum level two ingredients are necessary: spatial confinement of the polaritons and pronounced nonlinearity. The first can be achieved by etching of a planar microcavity, which allows to get individual polariton pillars \cite{Bajoni2008_PRL,Ctistis2010,Ferrier2011,Real2021,Sedov2021,Kuriakose2022}, systems of several coupled pillars forming so-called polariton molecules \cite{Galbiati2012} or periodically arranged arrays of the pillars forming polariton superlattices \cite{Milicevic2017,Suchomel2018,Whittaker2018,Whittaker2021}. As for polariton nonlinearity, it is provided by the excitonic fraction, and the main contribution is given by the exchange interaction of electrons and holes forming the excitons \cite{Ciuti1998,Glazov2009,Rochat2000_PRB,Shahnazaryan2016,Shahnazaryan2017}. The quantitative analysis of the polariton nonlinearities in micropillars is thus an important problem, which we solve in the current work.

    We consider in detail the formation of exciton-polaritons in cylindric micropillars, develop a microscopic theory for the calculation of the matrix elements of polariton-polariton interaction in this geometry, and analyze nonlinear dynamics of such a system subject to a partially coherent pumping, investigating, in particular, the redistribution of energy between polaritonic states with different angular momenta.

    The paper is organised as follows. In sections~\ref{sec:Photonic_modes}, \ref{sec:Excitonic_modes}, and \ref{sec:Polaritonic_modes} we briefly describe the formation of the photonic cavity modes, quantum-well excitonic modes (accounting for exciton-exciton interaction), and polaritonic modes, respectively. In section~\ref{sec:Dynamics} we consider the dynamics of the coherently pumped polaritons. Section~\ref{sec:RD} contains numeric results and their discussion. Conclusions are summarized in section~\ref{sec:Conclusion}. Appendices contain the technical details of the calculations.
  \end{section}

  \begin{section}{Theory \label{sec:Theory}}
    The system we consider is schematically shown in Figure~\ref{fig:micropillar} (left panel). It consists of a semiconductor micropillar with cylindrical symmetry (grey cylinder) flanked by a pair of dielectric Bragg mirrors (DBRs, the orange planes), and containing a quantum well (QW, the green circle in the middle) placed in the position of an antinode of an electric field of the optical cavity mode. The energy of the cavity mode is brought close to the resonance to the energy of the excitonic transition, favoring the formation of the hybrid polariton modes.

    As polaritons contain an excitonic fraction, they interact with each other. The scheme of the polariton-polariton interaction is shown in the right panel of Figure~\ref{fig:micropillar}. As the system we consider has cylindrical symmetry, polaritons are characterized by their projections of angular momenta ($\lambda_j$) onto the structure growth axis. States with different angular momenta have different profiles of their wave functions, shown schematically by the purple surfaces. Polaritons can exchange their angular momenta due to the exciton-exciton scattering, which, obviously, conserve the total angular momentum.

    The Hamiltonian of the interacting excitons coupled to a photonic cavity mode is:
    \begin{equation}
      \ham = \hx + \hp + \hpx + \hxx,
      \label{eq:Hamtot}
    \end{equation}
    where, $\hx$ and $\hp$ are Hamiltonians of free excitons and photons, $\hpx$ and $\hxx$ are Hamiltonians of photon-exciton and exciton-exciton interaction, respectively.

    \begin{figure}[t]
      \includegraphics[width=\columnwidth]{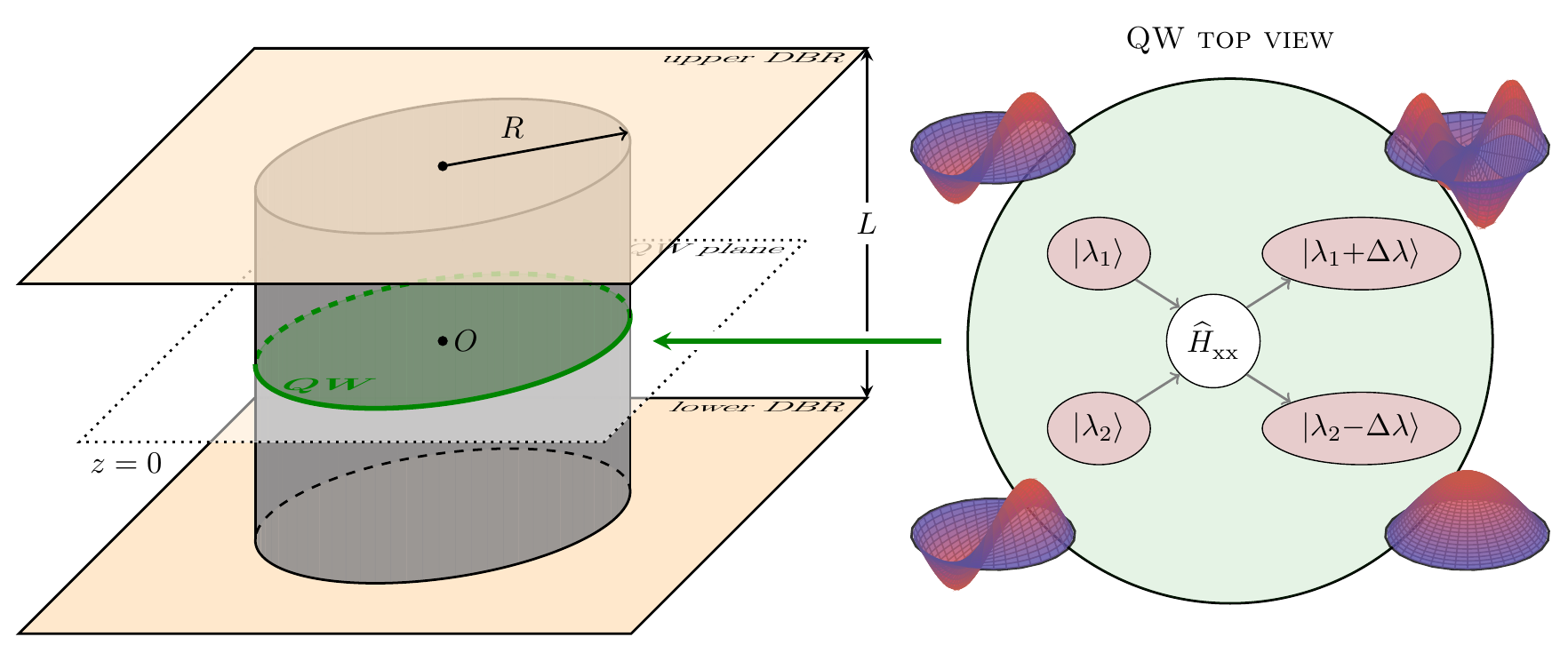}
      \caption{\label{fig:micropillar}  Left panel: Sketch of the considered geometry, consisting of a semiconductor micropillar (grey cylinder) flanked by a pair of dielectric Bragg mirrors (DBRs, the orange planes), and containing a quantum well (QW, the green circle in the middle) placed in the position of an antinode of an electric field of the optical cavity mode. The energy of the cavity mode is brought close to the resonance to the energy of the excitonic transition, giving rise to the formation of the  polaritons. Right panel: Top view of the system illustrating the scheme of the exciton-exciton scattering. Excitons in the states $|\lambda_1\rangle$, $|\lambda_2\rangle$, $|\lambda_1{+}\Delta\lambda\rangle$, and $|\lambda_2{-}\Delta\lambda\rangle$ (the red rounds) are coupled via interaction Hamiltonian $\hxx$ (the white round).
      Excitonic wave functions  $\lambda_1{=}\lambda_2{=}\Delta\lambda{=}1$ and $\nu_{1, 2, 3, 4} {=} 1$  are schematically shown by purple surfaces next to the states markers.}
    \end{figure}
    In second quantization representation, they can be written as:
    \begin{subequations}
      \begin{align}
        \hx &= \sumlmnu \hOmegaln \, \xcr \xan =
        \sum\limits_{\kappa} \hOmega_{\kappa} \, \x_{\kappa}^{\dagger} \x_{\kappa},
        \label{eq:Hp} \\
        \hp &= \sumln \homegaln \, \phcr \phan = \sum\limits_{k} \homega_k \, \ph_{k}^{\dagger} \ph_{k},
        \label{eq:Hx} \\
        \hpx &= \sum\limits_{k, \kappa} \hbar \parenth{g_{k \kappa} \x_{\kappa}^{\dagger} \ph_{k} + \hc},
        \label{eq:hamint} \\
        \hxx &= \frac{1}{2} \sum\limits_{\kappall} \vll \, \operator{\xsymbol}{\kappa_1}{\dagger} \, \operator{\xsymbol}{\kappa_2}{\dagger} \, \operator{\xsymbol}{\kappa_3}{} \, \operator{\xsymbol}{\kappa_4}{}.
        \label{eq:Hxx_int}
      \end{align}
    \end{subequations}
    Here, $\operator{\xsymbol}{\lambda\nu}{(\dagger)}$ and $\operator{\phsymbol}{ln}{(\dagger)}$ are the $\lambda\nu$-excitonic and $ln$-photonic mode annihilation (creation) operators, respectively, and $\hOmegaln$ and $\homegaln$ are the corresponding eigenenergies.
    The indices $\lambda$ and $\nu$ ($l$ and $n$) are the excitonic (photonic) angular and radial quantum numbers, respectively. In equation~\eqref{eq:hamint}, $\hbar g_{k\kappa}{\equiv}\hbar\gllnn$ represent the amplitudes of the interaction between $\lambda\nu$-excitonic and $ln$-photonic modes. To spare the notations, we hereafter denote the pairs of the excitonic, $\lambda_{(j)}\nu_{(j)}$, and photonic, $l_{(j)} n_{(j)}$, indices as $\kappa_{(j)}$ and $k_{(j)}$, respectively.

    \begin{subsection}{Photonic modes \label{sec:Photonic_modes}}

      To analyze the photonic modes in the cylindric symmetry microcavity, we follow the scheme of the reference \onlinecite{Snitzer1961_JOSA}.
      We consider a microcavity placed between a pair of DBRs and containing a cylindric micropillar of a radius $R$ made of a semiconductor with high-frequency dielectric permittivity $\epsilon_1$. The effective inter-DBRs distance is denoted as $L$. Within the conventional cylindric coordinates frame ($r$, $z$, and $\theta$), the $z$-component of the field is sought as a product of Bessel's functions (BFs) of either the first kind (core region, argument $\beta_1 r$) or the modified BF (air region, argument $\beta_2 r$), the complex exponential of $(i l\theta)$, and $\cos{\kz z}$. This introduces the radial, angular, and $z$ dependences, respectively. The value $\kz$ is chosen to be $\pi/L$ which corresponds to the fundamental (in $z$-direction) photonic mode and obeys the zero boundary conditions at the cylinder bases ($z = {\pm} L/2$).

      \begin{subequations}{\label{eq:Omega_characteristic_eq}}
        The use of the continuity condition for the fields tangential components at the core-to-vacuum boundary ($r=R$) leads to the characteristic equation (see \eqref{eq:characteristic_omegaphot}) and the photonic target function (PTF):
        \begin{align}
          \ptf (\hbar\omega) &= u^4 v^4 (\eta_1 + \eta_2) \parenth{k_1^2 \eta_1 + k_2^2 \eta_2} \nonumber \\
          &- l^2 \kz^2 \parenth{u^2 + v^2}^2,
          \label{eq:PTF}
        \end{align}
        where
        \begin{align}
          \eta_1 = \frac{J_l'(u)}{u \, J_l (u)} , \
          \eta_2 = \frac{K_l'(v)}{v \, K_l (v)} , \
          u = \beta_1 R, \
          v = \beta_2 R.
          \label{eq:uv_beta}
        \end{align}
        The PTFs are displayed in Fig.~\ref{fig:TFs_Hybr_diagram} as vertically oriented curves. Their roots (photonic modes eigenenergies) are marked by the round grey symbols.

        The core-region (index 1), air-region (index 2) quasimomenta $\beta_{1,2}$, and frequency $\omega$ are interconnected by
        \begin{equation}
          \beta_1^2 = \frac{\omega^2 \, \e_1}{c^2} - \kz^2, \quad
          \beta_2^2 = \kz^2 - \frac{\omega^2 \, \e_2}{c^2}.
          \label{eq:betas}
        \end{equation}
      \end{subequations}
      The system of eqs. \eqref{eq:Omega_characteristic_eq} is invariant under the $l \to -l$ substitution and, hence, is solved only for $l \geqslant 0$. The procedure results in a set of frequencies numbered by the index $n$ for each $l$: $\omega_{ln}$, $n = 1, 2, \ldots N(\abs{l})$.
    \end{subsection}

    All the details of the computational procedure are presented in Appendix~\ref{sec:Appendix_PhModes}.

    \begin{subsection}{Excitonic modes \label{sec:Excitonic_modes}}
      Considering the excitonic part, we use the following assumptions:
      \begin{equation}
        R \gg \abohr \gg \auc,
      \end{equation}
      where $\abohr$ is the two-dimensional exciton Bohr radius, and $\auc$ is the characteristic size of the micropillar material unit cell. These conditions allows to neglect the internal structure of excitons when considering their confinement inside a pillar. Moreover, the excitonic density $n_\text{X}$ is supposed to be small enough to treat them as bosons, i.e. the following condition is satisfied:
      \begin{equation}
          n_X \abohr^2\ll1.
      \end{equation}
      The dielectric-to-vacuum boundary is assumed to be an infinite barrier.

      Inside the cavity, we can factorize the excitonic wave function as
      \begin{equation}{\label{eq:phi_ex}}
        \phi_{\kappa} (\vr, \vrho) = \Phi_{\lmnu}(\vr) \, \chi_{1s} (\vrho),
      \end{equation}
      where $\Phi_{\lmnu} (\vr)$ describes the exciton centre-of-mass motion, and $\chi_{1s} (\vrho)$ corresponds to the relative motion of an electron and a hole, and for conventional semiconductor materials can be well approximated by the two-dimensional hydrogen-like atom wave function (in this paper, we consider only the $1s$ excitonic ground state):
      \begin{subequations}
        \begin{align}{\label{eq:Phi_exc}}
          \Phi_{\lmnu} (\vr) &= \frac{J_{\lambda} (\alpha_{\lmnu} r) \, \exp{i\lambda\theta}}{\abs{J_{\abs{\lambda}+1}(x_{\lmnu})} \, R \sqrt{\pi}}, \\
          \chi_{1s} (\vrho) &= \frac{\exp{\parenth{-\sfrac{\rho}{\abohr}}}}{\abohr} \sqrt{\frac{2}{\pi}}.
        \end{align}
      \end{subequations}
      The quantum numbers $\lambda$ and $\nu$ are thus associated with the exciton centre-of-mass motion, $x_{\lmnu}$ is the $\nu$th root of BF $J_{\lambda}$, and $\alpha_{\lmnu} = x_{\lmnu} / R$.

      The excitonic energies are expressed as:
      \begin{align}
        \hOmegaln = \frac{\hbar^2}{2 \parenth{m_{\elec}^* + m_{\hole}^*}} \parenth{\frac{x_{\lmnu}}{R}}^2
        - \frac{2\hbar^2}{\mu\abohr^2}
         + \Eg
        \label{eq:omegax}
      \end{align}
      with $m_{\elec}^*$ and $m_{\hole}^*$ being the electron and hole effective masses, respectively, $\mu$ -- the effective reduced mass of an electron-hole pair, and $\Eg$ -- the semiconductor energy band gap. The second term in \eqref{eq:omegax} is the $1s$-state eigenenergy.

      For realistic cavity parameters, we got that the first term, responsible for the dimensional quantization energy of the motion of the center of the mass is orders of magnitude smaller, then last two terms, so we can neglect the excitonic dispersion and take the energies of all excitonic states to be the same,
      \begin{equation}
        \hOmegaln = \hOmegax = \const,
      \end{equation}
      which is in agreement with the  reference~\onlinecite{Panzarini1999_PRB}.

      \begin{subequations}
       The matrix element of the exciton-exciton scattering $| \kappa_3, \kappa_4 \rangle \to | \kappa_1, \kappa_2 \rangle$ in the Born approximation can be written as
        \begin{equation}
          \vll = \langle \kappa_1, \kappa_2 | \vintxx | \kappa_3, \kappa_4 \rangle,
          \label{eq:Vkappas_as_<>}
        \end{equation}
        where $| \kappa_i, \kappa_j \rangle$ is the two-exciton wave function antisymmetric with respect the electron-electron and hole-hole exchange:
        \begin{gather}
          \nonumber
          | \kappa_i, \kappa_j \rangle = \frac12 \cdot
          \biggl[
          \phi_{\kappa_i} (\elec1, \hole1) \phi_{\kappa_j} (\elec2, \hole2) +
          \phi_{\kappa_i} (\elec2, \hole2) \phi_{\kappa_j} (\elec1, \hole1) \\
          -
          \phi_{\kappa_i} (\elec1, \hole2) \phi_{\kappa_j} (\elec2, \hole1) -
          \phi_{\kappa_i} (\elec2, \hole1) \phi_{\kappa_j} (\elec1, \hole2)
          \biggr]
        \end{gather}
        with $\elec 1(2)$ and $\hole 1(2)$ presenting the electron and hole coordinates, respectively.

        Within the same notations, the matrix elements of the exciton-exciton interaction potential, $\vintxx$, can be presented as\cite{Ciuti1998_PRB}
        \begin{equation}
          \vintxx = \vintc(\elec1, \hole2) - \vintc(\elec1, \elec2) + \vintc(\elec2, \hole1) - \vintc(\hole1, \hole2).
          \label{eq:Vintxx_as_sum}
        \end{equation}
        Here, $\vintc (\vr_1, \vr_2) = - e^2 / 4\pi\epsilon_0\estatic |\vr_1{-}\vr_2|$ is Coulomb potential, $\estatic$ is the micropillar static dielectric permittivity, $e$ is the elementary charge.
        The matrix element \eqref{eq:Vkappas_as_<>} can be expressed as the sum $\vll^{\dir} + \vll^{\xx} + \vll^{\text{ee}} + \vll^{\text{hh}}$ that reveals the four channels (direct, excitons exchange, electron-electron, and hole-hole exchange ones, respectively).

        The two latter terms jointly form the fermion-fermion exchange channel, $\vll^{\fx}$, which gives the major contribution \cite{Ciuti1998_PRB} and is only retained in our further consideration. The corresponding interaction matrix element is
        \begin{gather}
          \nonumber
          \vll = - \idotsint \dtau \, \phi_{\kappa_1}^* (\elec1, \hole1) \phi_{\kappa_2}^*(\elec2, \hole2) \, \vintxx \\
          \times \biggl[
          \phi_{\kappa_3} (\elec1, \hole2) \phi_{\kappa_4} (\elec2, \hole1) +
          \phi_{\kappa_3} (\elec2, \hole1) \phi_{\kappa_4} (\elec1, \hole2)
          \biggr],
          \label{eq:V_kappas_int8}
        \end{gather}
      \end{subequations}
      where $\dtau$ denotes the coordinate phase space volume differential.
      The elements $\vll^{\dir,\xx,\fx}$ respect the angular momentum conservation law, i.e. $\vll^{\dir,\xx,\fx}{=}0$ if $\lambda_1{+}\lambda_2 \not= \lambda_3{+}l_4$, which can be schematically written down in terms of the transferred angular momentum $\Delta \lambda$ in the scattering channel:
      \begin{equation}
        | \lambda_1 \rangle + | \lambda_2 \rangle \leftrightarrow | \lambda_1 + \Delta \lambda \rangle + | \lambda_2 - \Delta \lambda \rangle.
        \label{eq:XX_scat_Delta_lambda}
      \end{equation}
        \end{subsection}

    \begin{subsection}{The Hamiltonian in the polariton basis \label{sec:Polaritonic_modes}}
      \begin{figure*}[t]
        \includegraphics[width=\textwidth]{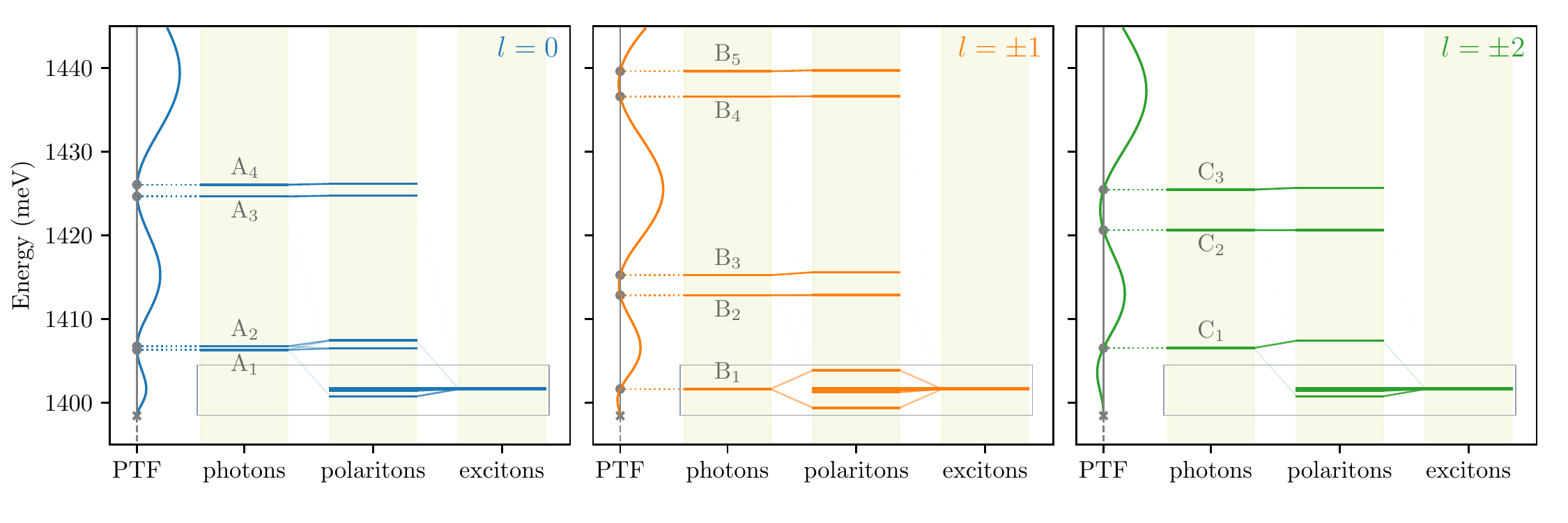}
        \caption{\label{fig:TFs_Hybr_diagram} The photonic target functions \eqref{eq:PTF} and hybridization schemes of exciton-polaritons formation for $l = 0, \pm 1, \pm 2$ angular quantum numbers. The vertical line at each PTF part shows the zero-PTF axis, the grey rounds and x-markers~-- the PTFs roots and lower photonic cut-off energies, respectively, the horizontal lines~-- the energy levels of photons (left), excitons (right), and polaritons (mid). The slant connection lines transparencies indicate the corresponding Hopfield coefficients (HCs) absolute values (the lower the value of the HC, the more the transparency). The open grey rectangles highlight the areas shown in Fig.~\ref{fig:Hybr_diagram_zoomed}.
        The micropillar is made of GaAs and has the radius $R = 1.5~\mum$, its effective height ($134.5~\nm$). The 2D Rabi splitting is $3.2~\meV$.}
      \end{figure*}
      Within the dipole approximation, the operator of exciton-photon interaction can be put down as
      \begin{equation}{\label{eq:ExPhot_int}}
        \vintln = - (\vdcv \cdot \vec{E}_{ln}),
      \end{equation}
      where $ln$ denotes the photonic mode, $\vdcv$ is the dipolar matrix element between valence and conduction bands. For the case of GaAs material (studied here), $\vdcv$ can be expressed as\cite{Ivchenko2005_AlphaSc}
      \begin{equation}
        \vdcv = \frac{\dcv}{\sqrt{2}} (\vec{e}_{x} + i \vec{e}_{y}) =
        \frac{\dcv}{\sqrt{2}} (\erho + i\etheta) \, \ee^{i\theta} =
        \vdcv^0 \, \ee^{i\theta}.
        \label{eq:dcv_er_etheta}
      \end{equation}

      Then, the exciton-photon interaction matrix element can be presented as the excitonic radiative linewidth multiplied by the overlap of the excitonic and photonic modes wave functions. Following the deduction of reference \onlinecite{Panzarini1999_PRB}, we eliminate the extra phase factor $\ee^{i \theta}$ in \eqref{eq:dcv_er_etheta} and arrive at the following expression for the interaction term:
      \begin{align}
        \hbar \gllnn &= - \chi_{1s}^* (0) \iint\limits_{\text{QW}} \Phi_{\lmnu}^* (\vr) \bigl[ \vdcv^0 (\vr) \cdot \vec{E}_{ln} (\vr) \bigr]_{z=0} \diff^2 \vr
        \label{eq:hg_llnn_as_int}
      \end{align}

      Utilizing the expression for the planar 2D microcavity Rabi splitting ($\hOmegaRS$, see Appendix~\ref{sec:Appendix_ExPh_coupling}), we recast the matrix element as
      \begin{equation}
        \hbar \gllnn = - \frac{\hOmegaRS}{2} \sqrt{\frac{\e_0 \e_1 L}{2\hOmegax}} \iint\limits_{\text{QW}} \Phi_{\lmnu}^* (\vr) \, E_{ln,||} (\vr) \, \diff^2 \vr,
     \end{equation}
where  $E_{ln,||} (\vr) = \bigl[ E_{ln, r}(\vr) + iE_{ln, \theta}(\vr) \bigr]_{z=0}$ is the sum of the in-plane radial and angular projections of the cavity mode electric field (with the latter projection being multiplied by imaginary unit). Note, that due to the rotational symmetry  only the excitons and photons possessing the same angular momentum ($l {=} \lambda$) are coupled, so that $\hbar \gllnn=0$ for $l\neq\lambda$. The corresponding explicit expressions are presented in Appendix~\ref{sec:Appendix_ExPh_coupling}.

The linear part of the Hamiltonian of the system in the exciton-photon basis can be diagonalized by means of a unitary transformation, introducing polariton modes $\plcr {\equiv} \pl_{\varkappa}^{\dagger}$ as linear combination of excitons and photons. The total Hamiltonian in the polariton basis reads:
      \begin{align}
        \ham &= \suml \left\{
        \sumn \homegaln \, \phcr \phan +
        \sumnu \hbar\Omega_{l\nu} \, \operator{X}{l\nu}{\dagger} \operator{X}{l\nu}{}
        \right.
        \nonumber
        \\
        &+ \left.
        \sumnnu \hbar \parenth{\glnn \xcrl \phan + \hc}
        \right\}
        + \hxx
        \label{eq:Hamtot_as_PandX} \\
        &= \sum\limits_{\varkappa} \hbar\varepsilon_{\varkappa} \, \pl_{\varkappa}^{\dagger} \pl_{\varkappa} +
        \frac12 \sum\limits_{\mathrlap{\varkappall}} W_{\varkappall} \, \pl_{\varkappa_1}^{\dagger} \pl_{\varkappa_2}^{\dagger} \pl_{\varkappa_3} \pl_{\varkappa_4},
        \tag{\ref{eq:Hamtot_as_PandX}$^\prime$},
        \label{eq:Hamtot_as_Pl}
      \end{align}
      where we used the notations $|\varkappa\rangle \equiv |l(\varkappa), n(\varkappa) \rangle \equiv |l, n \rangle$ or simply $|l\rangle$ (where relevant) to label different  polaritonic states. All the mathematical details, together with explicit expressions connecting $V_{\kappall}$ with $W_{\varkappall}$ are provided in Appendix~\ref{sec:Appendix_Transition_to_Pkappa}.

    \end{subsection}

    \begin{subsection}{Polaritons dynamics \label{sec:Dynamics}}
      The system of interacting polaritons can serve as a playground to examine the nonlinear dynamics in the regime of coherent optical pump.

      In the mean-field approximation, the Heisenberg equation for the operators $\pl_{\varkappa}$ can be transformed into the equation of motion for the the coherent amplitudes $c_{\varkappa} = \langle \pl_{\varkappa} \rangle$ which reads:
      \begin{align}
        \nonumber
        \dot{c}_{\varkappa} &= -\parenth{i \varepsilon_{\varkappa} + \frac{\gamma}{2}} c_{\varkappa} - \frac{i \pump_{\varkappa}}{\hbar} \exp{(-i \varepsilon_{\varkappa}t)} \\
        &-\frac{i}{2\hbar} \sum W_{\varkappall} \parenth{\delta_{\varkappa_1, \varkappa} c_{\varkappa_2}^* + \delta_{\varkappa_2, \varkappa} c_{\varkappa_1}^*} c_{\varkappa_3} c_{\varkappa_4}.
        \label{eq:dot_c_dynamics}
      \end{align}
      Here, we have neglected quantum correlations and used the approximation $\langle \pl_{\varkappa_j}^{\dagger} \pl_{\varkappa_m} \pl_{\varkappa_s} \rangle=c_{\varkappa_j}^* c_{\varkappa_m} c_{\varkappa_s}$.\cite{Witthaut2017_NatComm}
      The variables $\pump_{\varkappa}$ and $\gamma$ indicate an amplitude of the $\varkappa$ state coherent pumping and polariton decay rate, respectively. The latter is supposed to be the same for all polaritonic states.

      An example of minimal set of states that can demonstrate, under the zero initial conditions, the redistribution of the polaritons between different states is
      \begin{equation}
        |1\rangle + |1\rangle \leftrightarrow |2\rangle + |0\rangle,
        \label{eq:11->20_channel}
      \end{equation}
      which corresponds to the channel with angular momentum transfer of $\Delta l = 1$.
    \end{subsection}
  \end{section}

  \begin{section}{Results and discussion \label{sec:RD}}
    For all the calculations, the following cavity parameters were accepted: the micropillar material is GaAs with high-frequency dielectric permittivity $\e_1 = 10.89$ and the static one $\estatic = 12.9$, the cylinder radius $R = 1.5~\mum$, the effective inter-DBRs distance $L = 134.5~\nm$, the 2D Rabi splitting (twice the Rabi energy) $\hOmegaRS= 3.2~\meV$, the polariton linewidth $\hbar\gamma = 70~\mu\eV$. These parameters correspond to the experimental configuration of the reference~\onlinecite{Pernet2022_NatPhys}. The effective micropillar height, $L$, has been varied to get the different values of the detuning between photonic and excitonic modes, and has been finally adjusted to match the resonant case when the lower photonic mode energy ($B_1$) equals the energy of $1s$ exciton. This condition resulted in $L = 134.5~\nm$.

    Figure~\ref{fig:TFs_Hybr_diagram} demonstrates both the dependencies of the photonic target function (PTF) on the energy (left parts at each panel) and the hybridization diagram of the exciton-polariton formation for $l = 0, \pm1, \pm2$. The numerically determined roots of the PTFs (which give the photonic eigenenergies) are shown by the grey rounds, the horizontal dashed lines explicitly connect them with the corresponding photon levels at the diagrams.
    The calculations have yielded the total amount of 10, 10, and 8 photonic modes for $\abs{l} = 0, 1$ and $2$, respectively (the higher photonic modes are not displayed in the Figure (the energy axis is truncated by the value $1450~\meV$ from the top).

    Fig.~\ref{fig:Hybr_diagram_zoomed} presents the zoomed in sections of Fig.~\ref{fig:TFs_Hybr_diagram} outlined by the grey rectangles and illustrates the scheme of polaritons formation in more detail.
    The states $\text{PA}_{1l}$, $\text{PB}_{1l}$, and $\text{PC}_{1l}$ are the lowest polaritonic states with $l = 0$, ${\pm}1$, and ${\pm}2$, respectively.
    The percent values display the Hopfield coefficients, e.g. the two $50\%$ marks near photonic state $\text{B}_1$ mean that the latter splits into polaritonic states $\text{PB}_{1u}$ and $\text{PB}_{1l}$ in equal proportions (the separation energy equals $4.52~\meV$ which is $1.4$ times greater than $\hOmegaRS$).

    Each photonic mode with a given angular momentum, $l$, generally couples with the whole infinite set of excitonic states with $\lambda{=}l$. This gives rise, firstly, to the infinite number of polaritons with the identical $l$, and secondly, to the infinite number of terms in the excitons-over-polaritons decomposition used to calculate the $\widetilde{W}_{\varkappall}$ via $\widetilde{V}_{\kappall}$. On the other hand, some of the HCs are negligibly small, which is used to truncate the infinite-length decompositions, see Appendix~\ref{sec:Appendix_Trunc_x_as_p} for details.

    \begin{figure}[t]
      \includegraphics[width=.99\columnwidth]{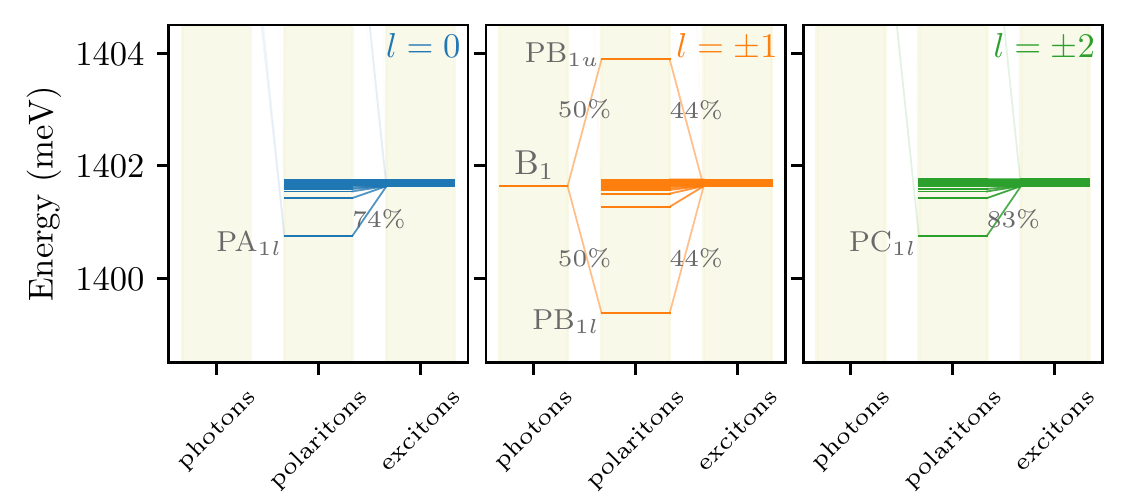}
      \caption{The lower polaritons formation scheme (for the states around the excitonic resonance, the magnified section of Fig.~\ref{fig:TFs_Hybr_diagram}). The states $\text{PA}_{1l}$, $\text{PB}_{1l}$, and $\text{PC}_{1l}$ are the lowest polaritonic states with $l = 0$, ${\pm}1$, and ${\pm}2$, respectively.
      The $\text{PB}_{1u}$ state is the counterpart to $\text{PB}_{1l}$ one: photonic mode $\text{B}_1$ splits into them almost equally. The percentage fractions present the absolute values of the corresponding Hopfield coefficients, squared. \label{fig:Hybr_diagram_zoomed}}
    \end{figure}

    \begin{figure*}[t]
      \includegraphics[width=.33\textwidth]{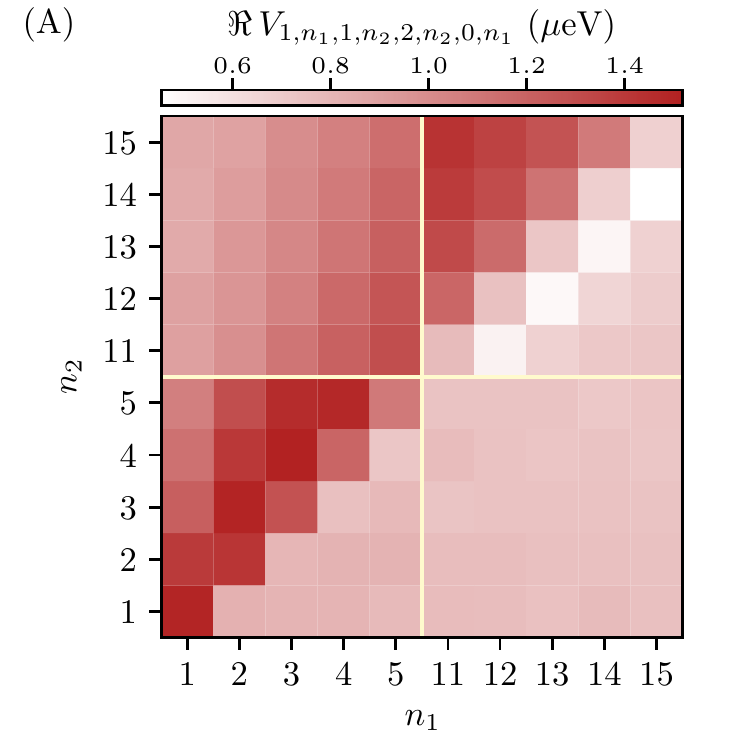}
      \hfill
      \includegraphics[width=.33\textwidth]{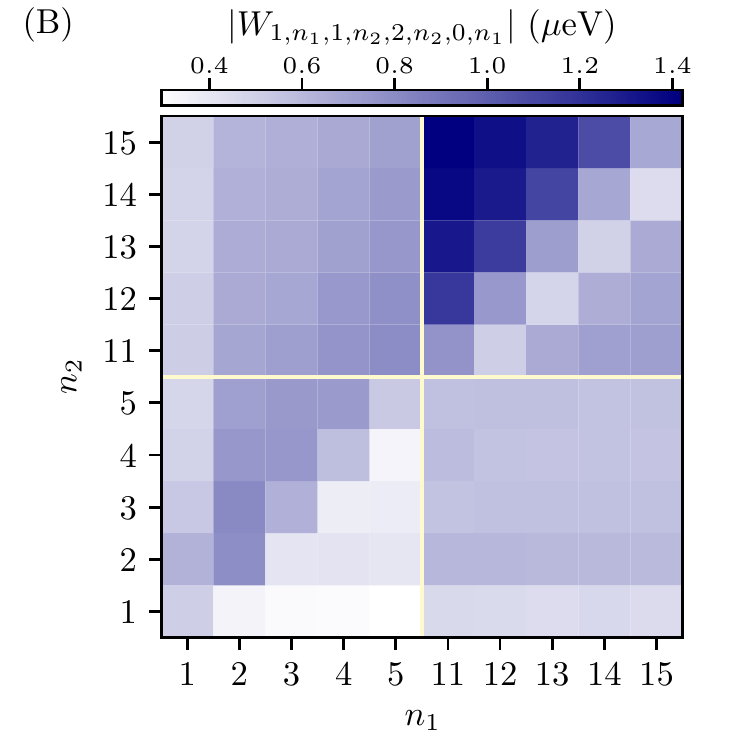}
      \hfill
      \includegraphics[width=.33\textwidth]{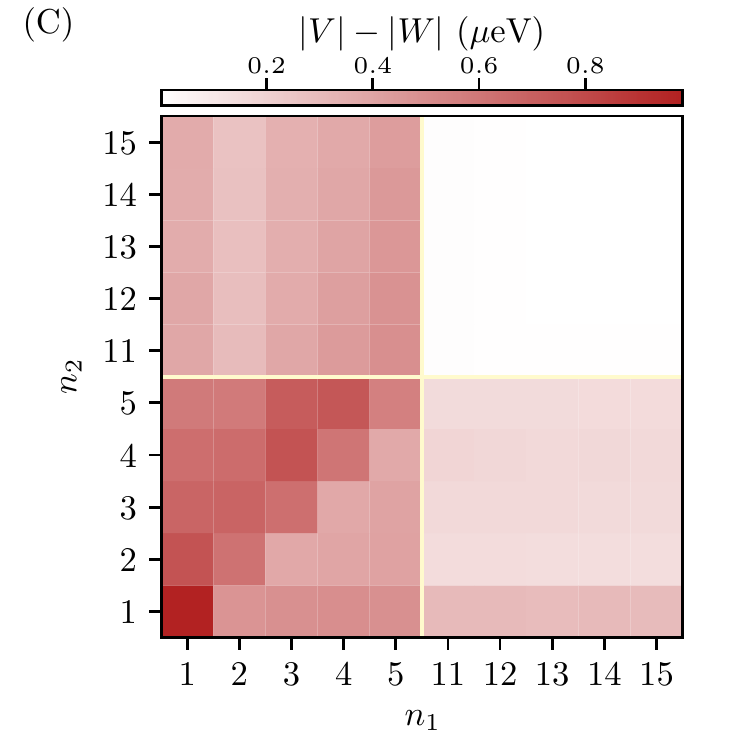}
      \caption{\label{fig:V_W_V-W} The absolute values of (A)~exciton-exciton, $V_{\kappall}$, and (B)~polariton-polariton, $W_{\varkappall}$, interaction matrix elements, and (C)~their difference, $\abs{V}{-}\abs{W}$. All the $V$s and $W$s are measured in $\mu\eV$. The states examined are $\kappa_1 {=} \varkappa_1 {=} (1, n_1)$, $\kappa_2 {=} \varkappa_2 {=} (1, n_2)$,
      $\kappa_3 {=} \varkappa_3 {=} (2, n_2)$, $\kappa_4 {=} \varkappa_4 {=} (0, n_1)$, with integers $n_1, n_2 = 1,\ldots 5, 11, \ldots 15$.
      The essence of the colour schemes used for all the panels is common: the associated absolute values grow with the increase in the colour saturation. The horizontal colour bars show the scales.}
      \vspace{.1in}
      \includegraphics[width=.24\textwidth]{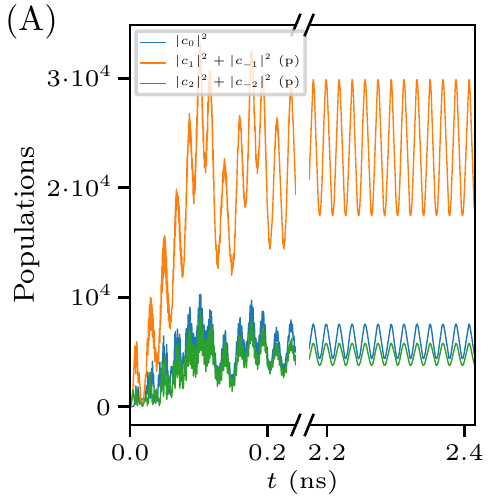} \hfill
      \includegraphics[width=.24\textwidth]{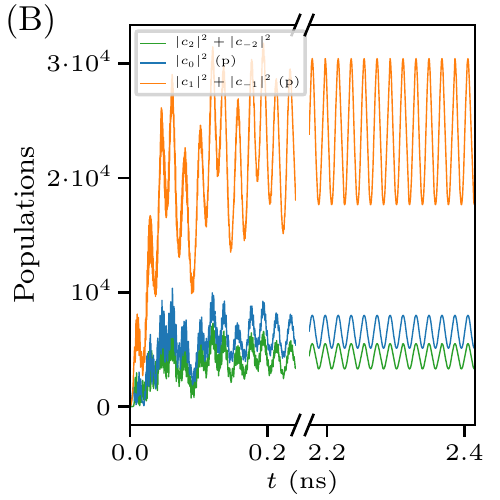} \hfill
      \includegraphics[width=.24\textwidth]{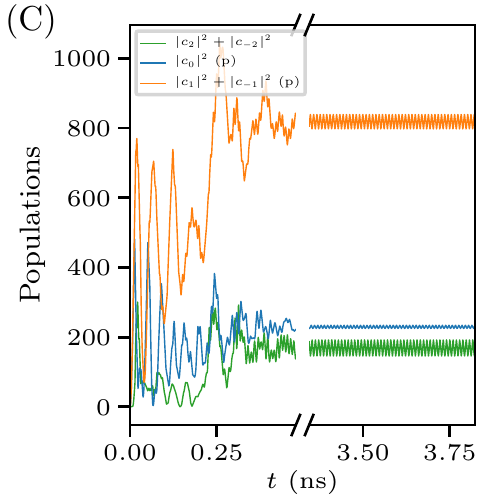} \hfill
      \includegraphics[width=.24\textwidth]{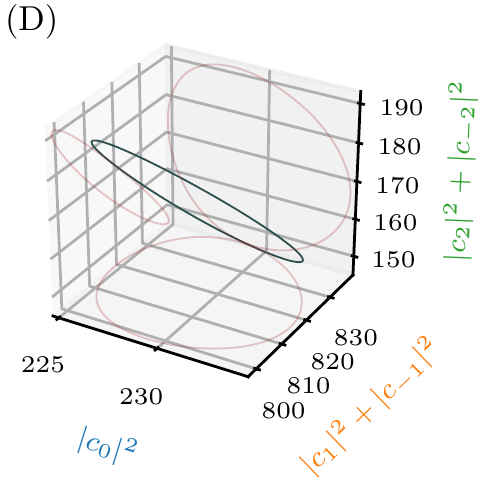}
      \caption{\label{fig:dynamics} (A, C) The populations dynamics of the system consistent of the five polaritonic states: $\abs{c_{0} (t)}^2$ (blue lines), $\abs{c_{1} (t)}^2 + \abs{c_{-1} (t)}^2$ (red lines), and $\abs{c_{2} (t)}^2 + \abs{c_{-2} (t)}^2$ (green lines), where the indices stand for the angular momenta. The lowest polaritonic modes ($n_l {=} 1$) are taken for all $l$.
      The time axis is broken to show both the transient period and the periodic behaviour.
      (B, D) The phase trajectories (the limit cycles here; black solid lines) for the later times (right parts in (A, C)).
      The states with (A, B) $l{=}\pm1, \pm2$, and (C, D) $l{=}0, \pm1$ are coherently pumped, which is marked by (p) in the legends. The semi-transparent red curves are the projections onto the coordinate planes.}
    \end{figure*}
    \textit{The exciton-exciton interaction coefficients} $V_{\kappall}$ are defined via the 8-dimensional integral \eqref{eq:V_kappas_int8} and have been computed by virtue of the Monte-Carlo integration algorithm (utilizing both the importance and uniform random samplings). The number of integrand evaluations was regulated by the maximal sufficient absolute error of $0.01~\mu\eV$. Also as $\vintc$ has a $1/r$ singularity, necessary changes of variables have been performed to integrate each summand of $\vintxx$ \eqref{eq:Vintxx_as_sum} in \eqref{eq:V_kappas_int8}.
    The real parts
    \begin{equation}
      \widetilde{V}_{n_1 n_2} \equiv \Re \, V_{1, n_1; 1, n_2; 2, n_2; 0, n_1}
    \end{equation}
    for $n_1, n_2 = 1, \ldots 5, 11, \ldots 15$ are presented in Figure~\ref{fig:V_W_V-W}(A), the numeric data are gathered in Table~\ref{eq:Appendix_V_n1n2}.

    \textit{The polariton-polariton interaction matrix elements}
    \begin{equation}
      \widetilde{W}_{n_1 n_2} \equiv \abs{W_{1, n_1; 1, n_2; 2, n_2; 0, n_1}}
    \end{equation}
    for the same set of $n_1 $ and $ n_2$ have been then evaluated via the sums \eqref{eq:W_as_sum_V}, and are presented in Figure~\ref{fig:V_W_V-W}(B) and Table~\ref{eq:Appendix_W_n1n2}.
    Figure~\ref{fig:V_W_V-W}(C) completes the picture with the difference $\abs{\widetilde{V}_{n_1 n_2}} - \abs{\widetilde{W}_{n_1 n_2}}$.

    Each of the panels in Fig.~\ref{fig:V_W_V-W} is divided into four quadrants in accordance with the groups of numbers $n_1$ and $n_2$. The transitions from panel (A) to (B) and (C) is quantitatively distinct for different quadrants. The ranges $(1\ldots 5)$ and $(11\ldots15)$ principally differ by the values of HCs performing the transition from $V_{\kappall}$ to $W_{\varkappall}$, see \eqref{eq:Appendix_HCs_nums} and \eqref{eq:Appendix_W=AV}: for the latter range the corresponding HCs absolute values are almost equal to unity.
    That is why $\abs{\widetilde{W}_{n_1, n_2}} \approx \abs{\widetilde{V}_{n_1 n_2}}$  when $n_1, n_2 \gtrsim 10$, which is clearly seen by the white coloured area in Fig.~\ref{fig:V_W_V-W}(C).

    \textit{The dynamics} of the micropillar polaritons \eqref{eq:dot_c_dynamics} was studied in more detail for the system consistent of five interacting polaritonic states subject to partial coherent pumping (PCP). The states are $|0 \rangle$, $|{\pm}1 \rangle$, and $|{\pm}2 \rangle$ with the fixed $n_j = 1$, their eigenenergies~--
    $\hbar\varepsilon_{0, 1} {=} 1400.752~\meV$, $\hbar\varepsilon_{\pm1, 1} {=} 1399.376~\meV$, and $\hbar\varepsilon_{\pm2, 1} {=} 1400.751~\meV$ (the data are put down with that precision as the interaction matrix elements are of $\mu\eV$ order of magnitude).
    The states are $\text{PA}_{1l}$, $\text{PB}_{1l}$, and $\text{PC}_{1l}$ in Fig.~\ref{fig:Hybr_diagram_zoomed}, their excitonic parts wave functions are the purple surfaces on the right panel of Fig.~\ref{fig:micropillar}.
    For the system considered, there are 85 nontrivial matrix elements $W_{\varkappall}$ that meet the angular momentum conservation law (see Appendix~\ref{sec:Appendix_W_dyn}), their absolute values vary from $0.35$ to $1.34~\mu\eV$.
    The model decay rate $\hbar\gamma = 70~\mu\eV$\cite{Pernet2022_NatPhys} corresponds to a lifetime $\tau \simeq 59~\text{ps}$.
    The zero initial conditions Cauchy problem was numerically integrated by means of the classic Runge--Kutta method with an invariant step of ca. $5~\text{fs}$.

    Figures~\ref{fig:dynamics}(A), (B), and (C) exhibit the dynamics of the states populations ($\abs{c_0 (t)}^2$, $\abs{c_1 (t)}^2+\abs{c_{-1} (t)}^2$, and $\abs{c_2 (t)}^2+\abs{c_{-2} (t)}^2$), panel (D) shows the phase trajectory of the dynamics (C).
    The states are coherently pumped under the following schemes:
    (1)~$\pump^{(1)}_{0} {=} 0$, $\pump^{(1)}_{1} {=} \pump^{(1)}_{-1} {=} \pump^{(1)}_{2} {=} \pump^{(1)}_{-2} {=} \pump/\sqrt{2}$ (panel A),
    (2)~$\pump^{(2)}_{0} = \pump$, $\pump^{(2)}_{1} {=} \pump^{(2)}_{-1} {=} \pump/\sqrt{2}$, $\pump^{(2)}_{\pm2} {=} 0$ (panel B), and
    (3)~$\pump^{(3)}_{0} = 0.2\pump$, $\pump^{(3)}_{1} {=} -\pump^{(3)}_{-1} {=} 0.2\pump/\sqrt{2}$, $\pump^{(3)}_{\pm2} {=} 0$ (panels C and D),
    with common notation $\pump = 40~\meV$.

    The results for the schemes (1) and (2) demonstrate several similar features: i)~the populations swing around close corresponding  values, ii)~both limit trajectories are limit cycles (not shown here) which are the line segments due to the phase synchronization of the vibrating populations, iii)~the long-range periods $\tau^{(1)} \approx \tau^{(2)} \approx 19~\text{ps}$, iv)~the states arranged in ascending order by the overall populations (at distant times) are $|{\pm}2\rangle$, $|0\rangle$, and $|{\pm}1\rangle$. On the other hand, this order can be treated as inverse for the scheme (1) as soon as state $|0\rangle$ is not directly pumped and, nevertheless, its population (averaged over the limit period) exceeds that of the pair of pumped states, $|{\pm}2\rangle$, see Fig.~\ref{fig:dynamics}(B).

    Scheme~(3) differs from (1) and (2) as it assumes the possibility to control the phase of the pumping field for the states with similar energies and distinct angular momenta. In the calculation, the pumping fields for $|{+}1\rangle$ and $|{-}1\rangle$ states have the same amplitudes and the phase shift of $\pi$. The limit cycle in the case acquires a pronounced oval shape.

    These results demonstrate, that the system can be regarded as a source of photons carrying angular momenta different from those which are directly excited by the coherent pumping.
  \end{section}

  \begin{section}{Conclusion \label{sec:Conclusion}}
   We considered theoretically the formation of exciton-polaritons in cylindric micropillars , their interaction, and pump-loss dynamics.   The exciton-exciton interaction matrix elements were numerically computed by means of the Monte-Carlo integration algorithm and converted to polariton-polariton ones through the unitary transformation. To analyze the nonlinear optical dynamics, we utilized the Heisenberg equation for the polaritonic operators and applied the mean-field truncation scheme which allowed us to study the effect of the nonlinear redistribution of the polariton between different confined modes. Our results can form a basis for the realization of a source of photons carrying non-zero orbital angular momenta.
  \end{section}

  \begin{acknowledgments}
    The work was supported by Priority 2030 Federal Academic Leadership Program. I.A.S. acknowledges support from the Icelandic Research Fund (Project ”Hybrid polaritonics”).
  \end{acknowledgments}

  \begin{appendix}
  \begin{widetext}

  \begin{section}{Cavity photonic modes \label{sec:Appendix_PhModes}}
    As described above, the microcavity is placed between the pair of DBRs and contains the cylindric micropillars of radius $R$ with the high-frequency dielectric permittivity $\epsilon_1$. The effective inter-DBRs distance is denoted as $L$.
    In this geometry, the $z$-components of the electric field inside the micropillar (core region, index 1) and outside it (air region, index 2) are sought in the form
    \begin{subequations}{\label{eq:fields_abcd}}
      \begin{align}
        E\inside_z (r, \theta, z, t) &= A_l \, J_l  (\beta_1 r) \, \cos{\kz z} \, \exp{i(l\theta - \omega t)}, \\
        E\outside_z (r, \theta, z, t) &= C_l \, K_l  (\beta_2 r) \, \cos{\kz z} \, \exp{i(l\theta - \omega t)},
      \end{align}
    \end{subequations}
    where $J_l$ and $K_l$ are $l$-th order Bessel's functions of the first and second kind, respectively, and $\kz {=} \pi/L$ gives the fundamental mode in $z$-direction.

    The condition of the fields tangential components continuity at the core-to-vacuum boundary ($r=R$) leads to the following characteristic equation:
    \begin{equation}{\label{eq:characteristic_omegaphot}}
      l^2 \cdot \kz^2 \cdot \parenth{\frac{1}{u^2} + \frac{1}{v^2}}^2 {=} (\eta_1 + \eta_2) \parenth{k_1^2 \eta_1 + k_2^2 \eta_2},
    \end{equation}
    and the corresponding target function presented in the main text.

    The frequencies $\omega_{ln}$ being determined, one then gets the electric field projections in the core region:
      \begin{subequations}{\label{eq:E_zrtheta}}
        \begin{align}
          E_{ln, z} (\vr) &= \cos{\kz z} \cdot \ee^{il\theta} \cdot A_{ln} \cdot J_l (x)
          , \\
          E_{ln, r} (\vr) &= \cos{\kz z} \cdot \ee^{il\theta} \cdot A_{ln} \cdot \frac{i\kz}{\beta_1} \cdot \brackets{J'_l (x) - l^2 \alpha_{ln} \frac{J_l (x)}{x}}, \\
          E_{ln, \theta} (\vr) &= \cos{\kz z} \cdot \ee^{il\theta} \cdot A_{ln} \cdot \frac{\kz l}{\beta_1} \cdot \brackets{J'_l (x) \, \alpha_{ln} - \frac{J_l (x)}{x}},
        \end{align}
      \end{subequations}
    where $x = \beta_1 r$, and
    \begin{equation}
      \alpha_{ln} = J_l (u) \cdot \parenth{\frac{1}{u^2} + \frac{1}{v^2}} \cdot \parenth{\frac{J_l'(u)}{u} + \frac{K_l'(v) \, J_l(u)}{K_l(v) \, v}}^{-1}.
      \label{eq:alpha}
    \end{equation}
    To get the temporal dependence, one is to multiply equations \eqref{eq:E_zrtheta} by $\ee^{-i\omega_{ln} t}$. To obtain the fields in the outer region, the amplitude $A_{ln}$ should be swapped for $C_{ln}$, $\beta_1$ for $\beta_2$, BFs of the first kind $J_l$ -- for the BFs of the second kind $K_l$, and their argument should be changed to $x {=} \beta_2 r$.
    The multiple subscript index $ln$ has been attached to $\alpha$ because of the implicit dependence of $u$ and $v$ on both $l$ and $n$ via $\beta_{1,2}$ \eqref{eq:betas}. The prefactors $A$ and $C$ (see eqs.~\eqref{eq:fields_abcd}) also acquire both indices: $A_{ln}$ and $C_{ln}$.

    To get the constant $A_{ln}$, we normalize the electric field by the energy of a single photon\cite{Gomes2020_JAP, Lifshitz1982_book}:
    \begin{align}
      \label{eq:E_normalization}
      \homegaln &= \epsilon_0 \iiint\limits_{\text{MC}} \epsilon(\vr) \, \abs{\vec{E}_{ln} (\vr)}^2 \, \diff^3 \vr = A_{ln}^2 \, G_{ln} \hence
      A_{ln} = \sqrt{\frac{\homegaln}{G_{ln}}}.
    \end{align}
    The integration is carried over the whole cavity volume, the dependence of the high-frequency permittivity on the position is denoted via $\epsilon(\vr)$ which equals $\epsilon_1$ inside the micropillar and unity outside.
    The normalization constant is calculated as
      \begin{multline}
        G_{ln} = \epsilon_0 \pi^2 \kz \cdot \left\{
        \frac{\varepsilon_1}{\beta_1^4}
        \int\limits_0^u x \, \diff x \brackets{(1 + l^2 \alpha^2_{ln}) \parenth{J'^2_l (x) + \frac{l^2 J_l^2 (x)}{x^2}} + \frac{\beta_1^2}{\kz^2} J_l^2 (x)}
        +
        \right. \\
        +
        \frac{1}{\beta_2^4}
        \int\limits_{v}^{+\infty} x \, \diff x \brackets{(1 + l^2 \alpha^2_{ln}) \parenth{K'^2_l (x) + \frac{l^2 K_l^2 (x)}{x^2}} + \frac{\beta_2^2}{\kz^2} K_l^2 (x)} \parenth{\frac{J_l(u)}{K_l(v)}}^2
        + \left. 2l^2 \alpha_l J_l^2 (u) \parenth{\frac{1}{\beta_2^4} - \frac{\varepsilon_1}{\beta_1^4}} \vphantom{\int\limits_{v}^{+\infty}} \right\}.
      \end{multline}
      The parameter $\alpha_{ln}$ is determined in \eqref{eq:alpha}. The integration is carried on over the quantum well ($0 \leqslant x \leqslant u$) and over the outer region ($x \geqslant v$).
  \end{section}

    \begin{section}{Exciton-photon coupling \label{sec:Appendix_ExPh_coupling}}
      Recall eq.~\eqref{eq:hg_llnn_as_int} for the exciton-photon interaction amplitude and extend it by inserting the expressions for the excitonic overlap wave function $\Phi_{\lmnu}$ \eqref{eq:Phi_exc} and the electric field \eqref{eq:E_zrtheta}, and integrating out the angular dependence $\ee^{i\theta (l-\lambda)}$:
      \begin{align}
        \hbar \gllnn &= - \chi_{1s} (0) \iint\limits_{\text{QW}} \Phi_{\lmnu} (\vr) \bigl[ \vdcv (\vr) \cdot \vec{E}_{ln} (\vr) \bigr]_{z=0} \diff^2 \vr {=}
        - \frac{\dcv \, \chi_{1s} (0)}{\sqrt{2}}
        \iint\limits_{\text{QW}} \Phi_{\lmnu} (\vr) \,
        \bigl[ E_{ln, r}(\vr) + i\, E_{ln, \theta}(\vr) \bigr]_{z=0} \, \diff^2 \vr \\
        &= - i \delta_{l\lambda}
        \dcv \chi_{1s} (0) \sqrt{\frac{2\homegaln}{G_{ln}}} \, \frac{\kz \, \pi}{\beta_1 \abs{J_{\abs{\lambda}+1}}  (x_\lmnu)}
        \int\limits_{0}^{R} r \, \diff r \, J_{\lambda} (\alpha_\lmnu r)
        \Biggl\{ \brackets{J'_l (x) - l^2 \alpha_{ln} \frac{J_l (x)}{x}} + l \brackets{J'_l (x) \, \alpha_{ln} - \frac{J_l (x)}{x}}
        \Biggr\} \\
        &\equiv \hbar \glnn \delta_{l\lambda},
      \end{align}
      where $x = \beta_1 r$.

      To make the theory uniform and utilize the experimental data for obtaining $\dcv$, we also derive the Rabi splitting for a square-shaped quantum well within the same formalism. In this case, the photonic field is given by
      \begin{equation}
        \vec{E}_{\vec{k}} (\vec{r}) = \ep \, A_{\vec{k}} \, \cos{\varkappa z} \, \ee^{i \vec{k} \vec{r}_{\parallel}},
      \end{equation}
      where $\varkappa$ carries the same meaning as earlier, $\vec{k}$ is the in-plane ($xy$) quasimomentum, $\ep$ is the polarization unit vector, $\vec{r}_{\parallel}$ is the in-plane radius-vector, $A_{\vec{k}}$ is the normalization constant which is defined via
      \begin{equation}
        \homega_{\vec{k}} = \e_0 \int\limits_{\text{MC}} \e(\vec{r}) \abs{\vec{E}_{\vec{k}} (\vec{r})}^2 \, \diff^3 \vec{r} =
        A_{\vec{k}}^2 \, \frac{\e_0 \e L S}{2} \hence
        A_{\vec{k}} = \sqrt{\frac{2 \homega_{\vec{k}}}{\e_0 \e LS}}.
      \end{equation}
      Here, $S$ is the lateral square of the rectangular shaped microcavity (coinciding with the QW square). The excitonic overlap function (for the centre of mass motion) is
      \begin{equation}
        \Phi_{\vec{k}} (\vec{r}_\parallel) = \frac{\ee^{i \vec{k} \vec{r}_{\parallel}}}{\sqrt{S}}
      \end{equation}
      with the same normalization area $S$.
      Therefore, the exciton-photon interaction amplitude is
      \begin{equation}
        \hbar g_{\vec{k} \vec{k}'} = -\chi_{1s} (0) \parenth{\vdcv \cdot \ep} \sqrt{\frac{2 \homega_{\vec{k}}}{\e_0 \e LS^2}}
        \iint\limits_{\text{QW}} \ee^{i \vec{r}_\parallel (\vec{k} - \vec{k}')} \, \diff x \, \diff y
        =
        -\chi_{1s} (0) \, \dcv \, \frac{\ep[x] + i \ep[y]}{\sqrt{2}} \, \sqrt{\frac{2 \homega_{\vec{k}}}{\e_0 \e L}} \, \delta_{\vec{k} \vec{k}'}.
        \label{eq:Appendix_hbarg_plane}
      \end{equation}
      The presence of $\delta_{\vec{k} \vec{k}'}$ indicates the coupling of the excitons and photons with the same in-plane quasimomenta.

      The Rabi splitting, $\hOmegaRS$, equals twice the amplitude (its absolute value) \eqref{eq:Appendix_hbarg_plane} for the case $\homega_{\vec{k}} = \hOmegax$:
      \begin{equation}
        \hOmegaRS = 2 \, \chi_{1s} (0) \, \dcv \, \sqrt{\frac{\hOmegax}{\e_0 \e L}}.
      \end{equation}
      Hence,
      \begin{equation}
        \chi_{1s} (0) \, \dcv = \frac{\hOmegaRS}{2} \, \sqrt{\frac{\e_0 \e L}{\hOmegax}},
      \end{equation}
      which is used in the main text, see Section~\ref{sec:Polaritonic_modes}.
    \end{section}

    \begin{section}{Transition to polaritonic basis \label{sec:Appendix_Transition_to_Pkappa}}
      Recall eqs. \eqref{eq:Hamtot_as_PandX} and \eqref{eq:Hamtot_as_Pl} for the total Hamiltonian:
      \begin{align*}
        \ham &= \suml \left\{
        \sumn \homegaln \, \phcr \phan +
        \sumnu \hbar\Omega_{l\nu} \, \operator{X}{l\nu}{\dagger} \operator{X}{l\nu}{}
        +
        \sumnnu \hbar \parenth{\glnn \xcrl \phan + \hc}
        \right\}
        + \hxx
        \tag{\ref{eq:Hamtot_as_PandX}} \\
        &= \sum\limits_{\varkappa} \hbar\varepsilon_{\varkappa} \, \pl_{\varkappa}^{\dagger} \pl_{\varkappa} +
        \frac12 \sum\limits_{\mathrlap{\varkappall}} W_{\varkappall} \, \pl_{\varkappa_1}^{\dagger} \pl_{\varkappa_2}^{\dagger} \pl_{\varkappa_3} \pl_{\varkappa_4}.
        \tag{\ref{eq:Hamtot_as_Pl}}
      \end{align*}
      The first three terms in \eqref{eq:Hamtot_as_PandX} form linear part of the total Hamiltonian. This part can be reduced to the diagonal form in the basis of polaritonic operators, $\plcr {\equiv} \pl_{\varkappa}^{\dagger}$, by a unitary transformation, thus giving the first sum in \eqref{eq:Hamtot_as_Pl}. As only the excitons and photons with same angular momenta are coupled, one can write this transformation in the form
      \begin{subequations}{\label{eq:BogoliuboW_transformation}}
        \begin{gather}
          \plan {=} \sum B_{l, n \nu} \operator{\xsymbol}{l \nu}{} +
          \sum C_{l, n n'} \operator{\phsymbol}{l n'}{}, \quad
          \operator{\xsymbol}{l \nu}{} {=} A_{l, \nu n} \plan
          \intertext{or shortly}
          \pl_{\varkappa} {=} \sum\limits_{\kappa} \Bex_{\varkappa\kappa} \x_{\kappa} +
          \sum\limits_{k} \Cph_{\varkappa k} \ph_{k},
          \quad
          \x_{\kappa} {=} \sum\limits_{\varkappa} \Aex_{\kappa \varkappa} \pl_{\varkappa},
          \label{eq:x_as_px}
        \end{gather}
      \end{subequations}
      keeping in mind that $l (\varkappa) {=} l (k) {=} \lambda (\kappa)$. Here, $\Bex_{\kappa k}$, $\Cph_{\varkappa k}$, and $\Aex_{\kappa\varkappa}$ are the 4-index matrices of the Hopfield coefficients of the direct ($\Bex$, $\Cph$) and inverse ($\Aex$) unitary transformations, respectively.

      The exciton-exciton interaction term, $\hxx$ \eqref{eq:Hxx_int}, can be rewritten in the polariton basis which leads to the polariton-polariton interaction:
      \begin{subequations}
        \label{eq:Hpp_and_VW}
        \begin{gather}
          \hxx \equiv \hpp= \frac12 \sum\limits_{\varkappall} W_{\varkappall} \, \pl_{\varkappa_1}^{\dagger} \pl_{\varkappa_2}^{\dagger} \pl_{\varkappa_3} \pl_{\varkappa_4},
          \label{eq:Hpp}
          \intertext{where}
          W_{\varkappall} {=} \sumprime\limits_{\mathclap{\kappall}}
          \Aex_{\kappa_1\varkappa_1} \Aex_{\kappa_2\varkappa_2} \Aex_{\kappa_3\varkappa_3} \Aex_{\kappa_4\varkappa_4}
          V_{\kappall}.
          \label{eq:W_as_sum_V}
        \end{gather}
      \end{subequations}
      is the corresponding matrix element. Formally, matrices $\Aex$ are of infinite size as there are infinite number of excitons possessing each angular momentum $\lambda$ (and hence there are the infinite number of polaritons for each $l$). Therefore, each sum in \eqref{eq:x_as_px} is formally of infinite length, which is overcome by a truncation, see Appendix~\ref{sec:Appendix_Trunc_x_as_p}.
    \end{section}

    \begin{section}{Exciton-exciton and polariton-polariton interaction matrix elements \label{sec:Appendix_VW}}
      In what follows, real parts of the exciton-exciton interaction matrix elements ($\tilde{V}_{n_1 n_2}$)
      \newcommand{\mathem}[1]{\mathbf{#1}}
      \begin{equation}
        \label{eq:Appendix_V_n1n2}
        \begin{array}{|c||c|c|c|c|c|c|c|c|c|c|}
          \hline
          \phantom{xx} 15 \phantom{xx} & \phantom{xx} 0.88 \phantom{xx} & \phantom{xx} 0.90 \phantom{xx} & \phantom{xx} 1.0 \phantom{xx} & \phantom{xx} 1.06 \phantom{xx} & \phantom{xx} 1.15 \phantom{xx} & \phantom{xx} \mathem{1.43} \phantom{xx} & \phantom{xx} \mathem{1.36} \phantom{xx} & \phantom{xx} \mathem{1.28} \phantom{xx} & \phantom{xx} \mathem{1.09} \phantom{xx} & \phantom{xx} \mathem{0.68} \phantom{xx} \\
          \hline
          14 & 0.86 & 0.92 & 1.02 & 1.09 & 1.20 & \mathem{1.40} & \mathem{1.31} & \mathem{1.12} & \mathem{0.69} & \mathem{0.45} \\
          \hline
          13 & 0.86 & 0.95 & 1.04 & 1.12 & 1.22 & \mathem{1.33} & \mathem{1.17} & \mathem{0.73} & \mathem{0.5} & \mathem{0.68} \\
          \hline
          12 & 0.91 & 0.96 & 1.06 & 1.18 & 1.27 & \mathem{1.19} & \mathem{0.76} & \mathem{0.49} & \mathem{0.66} & \mathem{0.70} \\
          \hline
          11 & 0.91 & 0.99 & 1.12 & 1.21 & 1.31 & \mathem{0.78} & \mathem{0.52} & \mathem{0.68} & \mathem{0.72} & \mathem{0.73} \\
          \hline
          5 & 1.07 & 1.31 & 1.47 & 1.49 & 1.10 & 0.74 & 0.74 & 0.74 & 0.72 & 0.73 \\
          \hline
          4 & 1.14 & 1.41 & 1.52 & 1.2 & 0.73 & 0.78 & 0.75 & 0.73 & 0.74 & 0.73 \\
          \hline
          3 & 1.23 & 1.51 & 1.29 & 0.76 & 0.79 & 0.74 & 0.75 & 0.75 & 0.75 & 0.74 \\
          \hline
          2 & 1.40 & 1.42 & 0.80 & 0.82 & 0.82 & 0.77 & 0.77 & 0.76 & 0.76 & 0.75 \\
          \hline
          1 & 1.50 & 0.83 & 0.82 & 0.81 & 0.79 & 0.78 & 0.77 & 0.75 & 0.78 & 0.76 \\
          \hline\hline
          \uparrow n_2 / n_1 \rightarrow & 1 & 2 & 3 & 4 & 5 & 11 & 12 & 13 & 14 & 15 \\
          \hline
        \end{array}
      \end{equation}
      and the polariton-polariton ones ($\tilde{W}_{n_1 n_2}$)
      \begin{equation}{\label{eq:Appendix_W_n1n2}}
        \begin{array}{|c||c|c|c|c|c|c|c|c|c|c|}
          \hline
          15 & \phantom{x} \phantom{-} 0.50 \phantom{x} & \phantom{x} -0.63 \phantom{x} & \phantom{x} \phantom{-} 0.65 \phantom{x} & \phantom{x} -0.68 \phantom{x} & \phantom{x} \phantom{-} 0.71 \phantom{x} & \phantom{x} \mathem{\phantom{-} 1.42} \phantom{x} & \phantom{x} \mathem{-1.35} \phantom{x} & \phantom{x-} \mathem{1.27} \phantom{x} & \phantom{x} \mathem{-1.09} \phantom{x} & \phantom{x} \mathem{\phantom{-} 0.68} \phantom{x} \\
          \hline
          14 & -0.49 & \phantom{-} 0.65 & -0.66 & \phantom{-} 0.70 & -0.74 & \mathem{-1.39} & \mathem{\phantom{-} 1.30} & \mathem{-1.12} & \mathem{\phantom{-} 0.68} & \mathem{-0.45} \\
          \hline
          13 & \phantom{-} 0.49 & -0.67 & \phantom{-} 0.67 & -0.71 & \phantom{-} 0.76 & \mathem{\phantom{-} 1.32} & \mathem{-1.16} & \mathem{\phantom{-} 0.73} & \mathem{-0.50} & \mathem{\phantom{-} 0.67} \\
          \hline
          12 & \phantom{-} 0.52 & -0.67 & \phantom{-} 0.68 & -0.75 & \phantom{-} 0.79 & \mathem{\phantom{-} 1.17} & \mathem{-0.75} & \mathem{\phantom{-} 0.49} & \mathem{-0.66} & \mathem{\phantom{-} 0.70} \\
          \hline
          11 & -0.52 & \phantom{-} 0.69 & -0.72 & \phantom{-} 0.77 & -0.81 & \mathem{-0.77} & \mathem{\phantom{-} 0.51} & \mathem{-0.67} & \mathem{\phantom{-} 0.72} & \mathem{-0.72} \\
          \hline
          5  & \phantom{-} 0.48 & -0.71 & \phantom{-} 0.75 & -0.74 & \phantom{-} 0.53 & \phantom{-} 0.58 & -0.58 & \phantom{-} 0.58 & -0.56 & \phantom{-} 0.57 \\
          \hline
          4  & \phantom{-} 0.50 & -0.76 & \phantom{-} 0.75 & -0.59 & \phantom{-} 0.35 & \phantom{-} 0.59 & -0.57 & \phantom{-} 0.56 & -0.57 & \phantom{-} 0.56 \\
          \hline
          3  & -0.54 & \phantom{-} 0.82 & -0.65 & \phantom{-} 0.38 & -0.38 & -0.57 & \phantom{-} 0.58 & -0.58 & \phantom{-} 0.58 & -0.58 \\
          \hline
          2  & \phantom{-} 0.64 & -0.80 & \phantom{-} 0.42 & -0.42 & \phantom{-} 0.41 & \phantom{-} 0.61 & -0.61 & \phantom{-} 0.61 & -0.61 & \phantom{-} 0.60 \\
          \hline
          1  & -0.51 & \phantom{-} 0.35 & -0.32 & \phantom{-} 0.31 & -0.30 & -0.47 & \phantom{-} 0.46 & -0.45 & \phantom{-} 0.47 & -0.46 \\
          \hline\hline
          \uparrow n_2 / n_1 \rightarrow & 1 & 2 & 3 & 4 & 5 & 11 & 12 & 13 & 14 & 15 \\
          \hline
        \end{array}
      \end{equation}
      are presented. All the values are given in $\mu\eV$. The imaginary parts are omitted as soon as their maximal absolute values are $0.01~\mu\eV$ and $0.001~\mu\eV$ for $\tilde{V}_{n_1 n_2}$ and $\tilde{W}_{n_1 n_2}$, respectively.
      The sections of \eqref{eq:Appendix_V_n1n2} and \eqref{eq:Appendix_W_n1n2} highlighted by the bold font are the ranges with close resemblance between the absolute values of $\tilde{V}_{n_1 n_2}$ and $\tilde{W}_{n_1 n_2}$. These areas form the upper-right quadrant in Fig.~\ref{fig:V_W_V-W}(C) which is almost white as the associated difference is vanishingly small. The values $\tilde{V}_{n_1 n_2}$ have been calculated with the accuracy of $0.01~\mu\eV$.
    \end{section}

    \begin{section}{Truncating the excitons-over-polaritons decompositions \label{sec:Appendix_Trunc_x_as_p}}
      The infinite-length decompositions of the excitons addressed in the main text and in Appendix~\ref{sec:Appendix_Transition_to_Pkappa} lead to each sum for $W_{\varkappall}$ \eqref{eq:W_as_sum_V} to be formally of the infinite length as well. To overcome it, we have, firstly, limited the number of excitonic modes (by 20 which is twice the maximal number of photonic modes with a fixed $\abs{l} \leqslant 2$).
      Secondly, we have truncated the decompositions of excitonic operators $\x_{\kappa}$ \eqref{eq:x_as_px} keeping only the summands with $\Aex_{\kappa \varkappa'}$ greater than half the achievable maxium in the row: if $\abs{\Aex_{\kappa \varkappa'}} \geqslant 0.5 \, \max\limits_{\varkappa} \abs{\Aex_{\kappa \varkappa}}$ with the condition that the maximum is not less than $0.1$ (actually, it was satisfied for all the states~-- there is no excitonic state with all the HCs less than $0.1$).

      For the excitonic states under the study, the operators can be schematically put down as
      \begin{equation}
        \label{eq:Appendix_HCs_nums}
        \arraycolsep=.5em
        \begin{array}{lllll}
        \operator{\xsymbol}{0, 1}{} \simeq \phantom{-} 0.86i \, \operator{\plsymbol}{0, 1}{}; &
        \operator{\xsymbol}{0, 2}{} \simeq \phantom{-} 0.82i \, \operator{\plsymbol}{0, 2}{}; &
        \operator{\xsymbol}{0, 3}{} \simeq \phantom{-} 0.77i \, \operator{\plsymbol}{0, 3}{}; &
        \operator{\xsymbol}{0, 4}{} \simeq -0.75i \, \operator{\plsymbol}{0, 4}{}; &
        \operator{\xsymbol}{0, 5}{} \simeq \phantom{-} 0.74i \, \operator{\plsymbol}{0, 5}{}; \\
        \operator{\xsymbol}{1, 1}{} \simeq \phantom{-} 0.66i \, \operator{\plsymbol}{1, 1}{} + 0.66i \, \operator{\plsymbol}{1, 21}{}; &
        \operator{\xsymbol}{1, 2}{} \simeq -0.85i \, \operator{\plsymbol}{1, 2}{}; &
        \operator{\xsymbol}{1, 3}{} \simeq \phantom{-} 0.84i \, \operator{\plsymbol}{1, 3}{}; &
        \operator{\xsymbol}{1, 4}{} \simeq \phantom{-} 0.85i \, \operator{\plsymbol}{1, 4}{}; &
        \operator{\xsymbol}{1, 5}{} \simeq \phantom{-} 0.84i \, \operator{\plsymbol}{1, 5}{}; \\
        \operator{\xsymbol}{2, 1}{} \simeq -0.91i \, \operator{\plsymbol}{2, 1}{}; &
        \operator{\xsymbol}{2, 2}{} \simeq -0.94i \, \operator{\plsymbol}{2, 2}{}; &
        \operator{\xsymbol}{2, 3}{} \simeq -0.92i \, \operator{\plsymbol}{2, 3}{}; &
        \operator{\xsymbol}{2, 4}{} \simeq \phantom{-} 0.90i \, \operator{\plsymbol}{2, 4}{}; &
        \operator{\xsymbol}{2, 5}{} \simeq \phantom{-} 0.93i \, \operator{\plsymbol}{2, 5}{}.
        \end{array}
      \end{equation}
      As seen, all the HCs are purely imaginary (which is valid for all other states as well). For $n \geqslant 11$, the absolute value of HC between $\operator{\xsymbol}{l, n}{}$ and $\operator{\plsymbol}{l, n}{}$ equals unity with the accuracy $\sim 10^{-2}$.

      Under the conditions, each of the sums \eqref{eq:W_as_sum_V} is reduced to a single term:
      \begin{equation}
        W_{\varkappall} =
        \Aex_{\varkappa_1 \varkappa_1} \Aex_{\varkappa_2 \varkappa_2}
        \Aex_{\varkappa_3 \varkappa_3} \Aex_{\varkappa_4 \varkappa_4}
        V_{\varkappall}.
        \label{eq:Appendix_W=AV}
      \end{equation}
      Due to the purely imaginary (for each $\Aex_{\varkappa \varkappa}$) and real (for $V_{\varkappall}$) structures, the matrix elements $W_{\varkappall}$ are also real, which is seen in \eqref{eq:Appendix_W_n1n2} and \eqref{eq:Appendix_W_l1l2l3l4}.
    \end{section}

    \begin{section}{$W_{\varkappall}$ matrix elements for the dynamics calculations \label{sec:Appendix_W_dyn}}
      As mentioned in the main text, there are totally $85$ nontrivial $W_{\varkappall}$ coefficients for scattering among $|0\rangle$, $|{\pm}1\rangle$ and $|{\pm}2\rangle$ polaritonic states (with all $n(\kappa_j){=}0$). For the sake of brevity, we shrink the indices $\varkappall$ to the lists $(l_1, l_2, l_3, l_4)$ and put down the $W$-values:
      \begin{align*}
        W_{0, 0, 0, 0} &= \phantom{-} 1.342~\mu\eV, \\
        W_{-2, 2, 2, -2} {=} W_{2, -2, 2, -2} {=} W_{-2, 2, -2, 2} {=} W_{2, -2, -2, 2} &= \phantom{-} 1.228~\mu\eV, \\
        W_{2, 2, 2, 2} {=} W_{-2, -2, -2, -2} &= \phantom{-} 1.226~\mu\eV, \\
        W_{0, -2, 0, -2} {=} W_{0, 2, 2, 0} {=} W_{2, 0, 2, 0} {=} W_{-2, 0, -2, 0} {=} W_{2, 0, 0, 2} {=} W_{-2, 0, 0, -2} {=} W_{0, 2, 0, 2} {=} W_{0, -2, -2, 0} &= \phantom{-} 0.769~\mu\eV, \\
        W_{0, 0, -2, 2} {=} W_{0, 0, 2, -2} &= \phantom{-} 0.768~\mu\eV, \\
        W_{2, -2, 0, 0} {=} W_{-2, 2, 0, 0} &= \phantom{-} 0.763~\mu\eV, \\
        W_{-1, -2, -2, -1} {=} W_{-1, -2, -1, -2} {=} W_{1, 2, 2, 1} {=} W_{-2, -1, -1, -2} {=} W_{1, 2, 1, 2} {=} W_{2, 1, 1, 2} {=} W_{2, 1, 2, 1} {=} W_{-2, -1, -2, -1} &= \phantom{-} 0.622~\mu\eV, \\
        W_{2, -2, -1, 1} {=} W_{-2, 2, -1, 1} {=} W_{2, -2, 1, -1} {=} W_{-2, 2, 1, -1} {=} W_{1, -1, 2, -2} {=} W_{-1, 1, -2, 2} {=} W_{-1, 1, 2, -2} {=} W_{1, -1, -2, 2} &= -0.613~\mu\eV, \\
        W_{1, -2, -2, 1} {=} W_{-1, 2, -1, 2} {=} W_{-1, 2, 2, -1} {=} W_{-2, 1, -2, 1} {=} W_{2, -1, -1, 2} {=} W_{1, -2, 1, -2} {=} W_{2, -1, 2, -1} {=} W_{-2, 1, 1, -2} &= \phantom{-} 0.612~\mu\eV, \\
        W_{0, 0, -1, 1} {=}  W_{0, 0, 1, -1} {=}  W_{1, -1, 0, 0} {=}  W_{-1, 1, 0, 0} &= -0.543~\mu\eV, \\
        W_{1, 0, 1, 0} {=}  W_{0, -1, -1, 0} {=}  W_{0, 1, 1, 0} {=}  W_{-1, 0, 0, -1} {=}  W_{-1, 0, -1, 0} {=}  W_{1, 0, 0, 1} {=}  W_{0, 1, 0, 1} {=}  W_{0, -1, 0, -1} &= \phantom{-} 0.539~\mu\eV, \\
        W_{-1, 0, -2, 1} {=} W_{-1, 0, 1, -2} {=} W_{0, 1, -1, 2} {=} W_{1, 0, 2, -1} {=} W_{0, 1, 2, -1} {=} W_{0, -1, -2, 1} {=} W_{0, -1, 1, -2} {=} W_{1, 0, -1, 2} &= \phantom{-} 0.521~\mu\eV, \\
        W_{1, 1, 2, 0} {=} W_{1, 1, 0, 2} {=} W_{-1, -1, -2, 0} {=} W_{-1, -1, 0, -2} &= -0.516~\mu\eV, \\
        W_{-1, 2, 0, 1} {=} W_{-1, 2, 1, 0} {=} W_{1, -2, -1, 0} {=} W_{-2, 1, -1, 0} {=} W_{1, -2, 0, -1} {=} W_{2, -1, 0, 1} {=} W_{-2, 1, 0, -1} {=} W_{2, -1, 1, 0} &= \phantom{-} 0.515~\mu\eV, \\
        W_{0, -2, -1, -1} {=} W_{0, 2, 1, 1} {=} W_{-2, 0, -1, -1} {=} W_{2, 0, 1, 1} &= -0.512~\mu\eV, \\
        W_{1, -1, 1, -1} {=} W_{1, -1, -1, 1} {=} W_{-1, 1, -1, 1} {=} W_{-1, 1, 1, -1} &= \phantom{-} 0.347~\mu\eV, \\
        W_{1, 1, 1, 1} {=} W_{-1, -1, -1, -1} &= \phantom{-} 0.346~\mu\eV.
        \numberthis
        \label{eq:Appendix_W_l1l2l3l4}
      \end{align*}
    \end{section}

  \end{widetext}

\end{appendix}

  \bibliographystyle{unsrt}
  \bibliography{bibitems}

\end{document}